\def\asr{{ Advances in Space Research}}
\def\pop{{ Phys. Plasma}}
\def\ag{{ Annales Geophysicae}}
\def\lrsp{{ Living Reviews in Solar Physics}}

\def\spose#1{\hbox to 0pt{#1\hss}}
\def\approxlt{\mathrel{\spose{\lower 3pt\hbox{$\sim$}}
        \raise 2.0pt\hbox{$<$}}}
\def\approxgt{\mathrel{\spose{\lower 3pt\hbox{$\sim$}}
        \raise 2.0pt\hbox{$>$}}}

\def\multleft#1{\hbox to size{\vbox {\halign {\lft{##}\cr #1}}\hfill}\par}
\def\multright#1{\hbox to size{\vbox {\halign {\rt{##}\cr #1}}\hfill}\par}

\def\boxit#1{\vbox{\hrule\hbox{\vrule\kern3pt\vbox{\kern3pt
          #1 \kern3pt}\kern3pt\vrule}\hrule}}

\def\km{{\rm\thinspace km}}

\def\s{{\rm\thinspace s}}

\def\kmps{\hbox{$\km\s^{-1}\,$}}

\documentclass[preprint]{aastex}
\usepackage{graphicx}
\usepackage{color}
\usepackage{amsmath}
\received{}
\accepted{}
\slugcomment{submitted to ApJL} 
\shorttitle{The Origin of Electron Core-Halo Features}
\shortauthors{Che et al.}
\begin{document}
\title{The Origin of Non-Maxwellian Solar Wind Electron Velocity Distribution Function: Connection to Nanoflares in the Solar Corona }

\author{H. Che \& M. L. Goldstein}
\affil{NASA/Goddard Space Flight Center, Greenbelt, MD, 20771, USA}

\begin{abstract} 
The formation of the observed core-halo feature in the solar wind electron velocity distribution function is a long-time puzzle. In this letter based on the current knowledge of nanoflares we show that the nanoflare-accelerated electron beams are likely to trigger a strong electron two-stream instability that generates kinetic Alfv\'en wave and whistler wave turbulence, as we demonstrated in a previous paper. We further show that the core-halo feature produced during the origin of kinetic turbulence is likely to originate in the inner corona and can be preserved as the solar wind escapes to space along open field lines. We formulate a set of equations to describe the heating processes observed in the simulation and show that the core-halo temperature ratio of the solar wind is insensitive to the initial conditions in the corona and is related to the core-halo density ratio of the solar wind and to the quasi-saturation property of the two-stream instability at the time when the exponential decay ends. This relation can be extended to the more general core-halo-strahl feature in the solar wind. The temperature ratio between the core and hot components is nearly independent of the heliospheric distance to the sun. We show that the core-halo relative drift previously reported is a relic of the fully saturated two stream instability. Our theoretical results are consistent with the observations while new tests for this model are provided.  
\end{abstract}

\keywords{sun: corona---Acceleration of particles---Instabilities---Turbulence---Scattering---Solar wind}

\section{introduction}
Even with decades of extensive studies, our understanding is still poor as to what physical processes produce the non-Maxwellian electron velocity distribution functions (EVDFs) in the solar wind \citep{feldman75jgr,pilipp87jgra}. Observations of EVDFs at heliocentric distances from 0.3 to 1 AU \citep{pilipp87jgrb} show a prominent ``break" or a sudden change of slope at a kinetic energy of a few tens of electron volts, suggestive of two electron populations: the ``core" that dominates below the break, and the ``halo" that dominates at higher energies. In addition, an anisotropic tail-like feature, skewed with respect to the magnetic field direction is found in the EVDF of solar wind with speed $>300$ \kmps. This is called the ``strahl".  In solar wind coming from a sector boundary with speed $<300$\kmps, the strahl is nearly invisible and the isotropic core-halo feature dominates. Kinetic studies suggest that the strahl can result from the trapping of electrons due to large-scale focusing of the interplanetary magnetic field even in the presence of Coulomb collisions and turbulent scattering
\citep{marsch06lrsp,vocks12ssr}. However, there is no model that produces the isotropic halo, nor one that predicts the properties of the observed core-halo feature.

\citet{lin97conf} discovered in the solar wind at 1 AU a previously unknown population of energetic electrons even during solar quiet-time, which he called  ``superhalo". It has energies of a few tens keV as was confirmed recently using data from the STEREO satellite \citep{wang12apjl}. The superhalo electrons probably arise from nanoflares in the solar corona. In a recent paper \citep[Paper I]{che14prl}, we postulated that nanoflares can produce electron beams with a broad range of energies. We extrapolate the superhalo density to the nanoflare-accelerated keV electrons assuming that the beam energy flux is a flat function of electron kinetic energy. We showed using a particle-in-cell (PIC) simulation that the beam electrons can release their kinetic energy via an electron two-stream instability that leads to production of kinetic turbulence in the solar wind. One of the important conclusions was the formation of an isotropic electron halo.

Given that the above model is largely based on solar wind observations and extrapolations of superhalo constraints, it is critically important to examine if nanoflare-accelerated electrons are sufficient to trigger the strong electron two-stream instability in the context of existing observational and theoretical knowledge of nanoflares. In this letter, we first show that our model assumptions are consistent with existing observations and theoretical estimates of the properties of nanoflares. We further show that the electron halo forms in the inner corona and can be persevered as the solar wind escapes along open field lines. Then we derive a set of equations that describe the heating mechanism discovered in paper I, and study how the heating mechanism determines the core-halo properties of the EVDF. We show that the solar wind core-halo temperature ratio is a function of the core-halo density ratio and the quasi-saturation condition of the two-stream instability, and is insensitive to the initial condition in the solar corona. The result can be extended to the more general core-halo-strahl feature. Our theoretical results agree well with observations. Moreover, the formation of the electron halo in the inner corona might have important implications for the models of field line launching and accelerating of solar wind \citep{scudder92apja,scudder92apjb,fisk03jgr,gloeckler03jgr,van10apj,he10asr}

 \section{Nanoflare-Accelerated Electrons and the Formation of Electron Halo}
\subsection{Formation of electron halo in the Inner Corona}
Small-scale bundled magnetic flux tubes rooted in the photosphere stretch into the solar corona. Photospheric convection leads to the shuffling of foot-points of the magnetic field, and produces magnetic reconnections that eject plasma from the photosphere into the corona along open magnetic field lines. Parker called these events ``nanoflares"~\citep{parker88apj}. Recent high spatial resolution X-ray observations of the sun are providing increasing evidence supporting the existence of nanoflares \citep{viall13book,viall13apj,win13apj,testa13apjl}. In particular, the High-resolution Coronal Imager carried on a NASA sounding rocket discovered a set of fast evolving magnetic loops in the inner-most region of the corona that had lifetimes, sizes and impulsive heating consistent with those expected for nanoflares~\citep{win13apj}. Typical energy released by nanoflares is $\sim 10^{24}$ ergs, $\sim 9$ orders of magnitude lower than that of flares~\citep{parker88apj}. The duration of nanoflares is typically a few seconds and the size of the loops is typically a few hundred kilometers, about one thousandth the size of flare loops \citep{aschw02apj}. The fundamental difference between nano- and normal flares is that nanoflares can continuously occur everywhere in the quiet sun, even in corona holes, while  flares only occur in active regions. The occurrence rate of nanoflares is nearly $10^6$ events/s over the whole sun, even during solar quiet-times. Similar to flares, nanoflares accelerate particles, and the characteristic energy of the accelerated electrons is in the keV range~\citep{gon13apj}. Therefore, nanoflares form a quasi-continuous source of free energy. 

 In Paper I we showed that Kinetic  Alfv\'en Wave (KAW) and whistler wave turbulence, as well as an electron halo, can be produced if the kinetic energy of electron beam is $\gtrsim 70$ times the thermal energy of the core, and the electron beam density is $\gtrsim10\%$ of the core electron density. We now examine if these conditions can be satisfied based on estimations using our current knowledge of nanoflares. We also show where the core-halo forms and whether the feature can survive collisions in the corona before escaping into space. 

Nearly 50\% of the released energy from solar flares is in the kinetic energy of accelerated particles \citep{lin11ssr}. The density of accelerated electrons can be comparable to the coronal density~\citep{krucker10apj}. Given the similarity between normal and nanoflares, for simplicity we can assume that all the energy released by nanoflares is converted into the electrons' kinetic energy of $\sim 1$~keV.  For the event rate of $\sim 10^{-17}$~cm$^{-2}$s$^{-1}$\citep{parker88apj,vekstein09aap}, the number of keV electrons produced by nanoflares at the corona base should be $n_{e,nano} \sim 6 \times 10^{15}$~cm$^{-2}$s$^{-1}$. We assume that these keV electrons are uniformly distributed in a thin layer above the surface of the sun with the thickness equals to the average height of nanoflare loops $\sim 1000$~km $\sim 0.01 R_{\bigodot}$\citep{feldman99apj,klim12jgr,chen_b13apjl}. Given that the typical duration of nanoflares is a few seconds, the density of keV electrons should be $\gtrsim 10^7$cm$^{-3}$. In the inner corona, the ambient electron density is $n_e \sim 10^8$ cm$^{-3}$ and hence $n_{e,nano}/n_e\gtrsim 10\%$. 
 
The observed electron temperature in the inner corona ranges from a few times $10^5$~K to $10^6$~K~\citep{esser95jgr}. The electron temperature of the ambient coronal gas before heating should be $\sim 10^5$~K. The kinetic energy of beam electrons is then $\sim 100$ times higher than the thermal energy of the coronal electrons before the onset of turbulent heating. Therefore, the conditions we adopted in Papar I is likely to be satisfied. 

In paper I, we showed that keV electrons can generate kinetic turbulence and produce an electron halo and become demagnetized and thermalized within $\sim 10$ ion gyro-periods or $\sim 10^{-4}$~s, assuming a magnetic field of 1G~\citep{gopa12apj}. The KeV electrons can travel $\sim 10$~km before thermalization, a distance much smaller than the height of typical magnetic loops. If the electrons are produced in the inner corona and travel along either magnetic loops or open field lines, the electrons are released to the corona and become the halo within $1.1R_{\bigodot}$. The electron Coulomb collision rate in the inner corona is $\nu_e \sim 10^{-6}n_eT_e^{-3/2}\sim 10^{-4}$~s$^{-1}$ and the Coulomb heating time scale is $10^4$ seconds. During this period the solar wind travels $\sim 10 R_{\bigodot}$ with a velocity of $\sim 100$~\kmps. Thus collisions are insufficient to scatter halo electrons into the core to form a single Maxwellian distribution before the expansion of corona convects the turbulent plasma into the outer corona at around $\sim 2-3 R_{\bigodot}$ where the electron density drops from $10^8~$cm$^{-3}$ to $10^4~$cm$^{-3}$~\citep{cranmer09lrsp} and can be neglected. Therefore, the core-halo shape is stable to collisional scattering as the solar wind being advected outward to 1~AU and beyond \citep{parker65ssr,jock70aap,fisk03jgr,feldman05jgr}.

\subsection{The Properties of EVDF }
To facilitate our discussion of the properties of EVDF, we first summarize the PIC simulation reported in Paper I. The 2.5D PIC simulation is initialized in the solar wind reference frame with a uniform magnetic field $B_0\hat{x}$. The initial EVDF has two Maxwellian components that represent the ambient thermal electrons in the corona and the nanoflare-accelerated keV electron stream, respectively. The ratio of the density of the electron beam $n_{h0}$ and that of the ambient electrons $n_{c0}$ is 0.1. The relative drift $v_{d0}$ between the two populations is along $B_0$. The ratio of the mean kinetic energy of beam electrons $E_{d0}=m_ev_{d0}^2/2$ and the thermal energy of ambient electrons $k_BT_{c0}$ is 72 ($k_B$ is the Boltzmann constant). 

\begin{figure}
\includegraphics[scale=0.7]{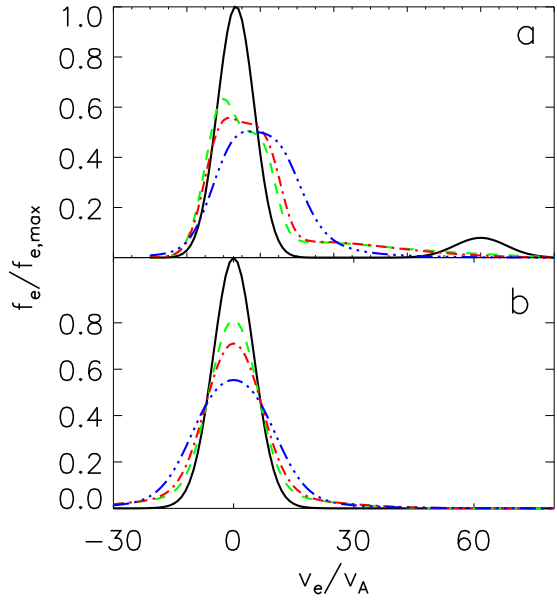}
\caption{ The EVDFs $f(v_x)=\int f(v_x, v_y)dv_y$ (panel a) and $f(v_y)=\int f(v_x, v_y)dv_x$ (panel b) at  $\Omega_{i} t=0$ (black solid line), 0.32, two-stream instability nearly saturates and slowly decay (green dashed), 0.64, the Weibel instability is near the peak (red dash-dotted), and 10.56, turbulence is in full saturation  (blue dots-dashed).
\label{dist}}
\end{figure}

We now look at the evolution of EVDF in the simulation. In Fig.~\ref{dist}, we show the EVDFs $f(v_x)$ and $f(v_y)$ at four times. At the beginning, two stream instability is triggered and quickly damps most of the kinetic energy of the electron beam, part of the beam electrons join the ambient electrons and the remaining form a long tail along $v_x$ (green dashed). This instability nearly saturates on the time scale of the electron plasma frequency $1/\omega_{pe}$. A Weibel-like instability is then triggered and converts the electrostatic waves into electromagnetic waves, which in turn scatters the beam electrons in direction perpendicular to the magnetic field and broadens the narrow tail. The breakup of the Weibel-like waves producing irregular wave-wave interactions further scatters the electrons into different directions. At the final stage the KAW and whistler wave turbulence evolve to an equilibrium state. Nearly all ($>$ 96\%) of the energy in electromagnetic fluctuations is returned to heat the electron tail. The tail is scattered into a nearly isotropic halo, resulting in a ``core-halo" structure.

\begin{figure}
\includegraphics[width=0.9\textwidth]{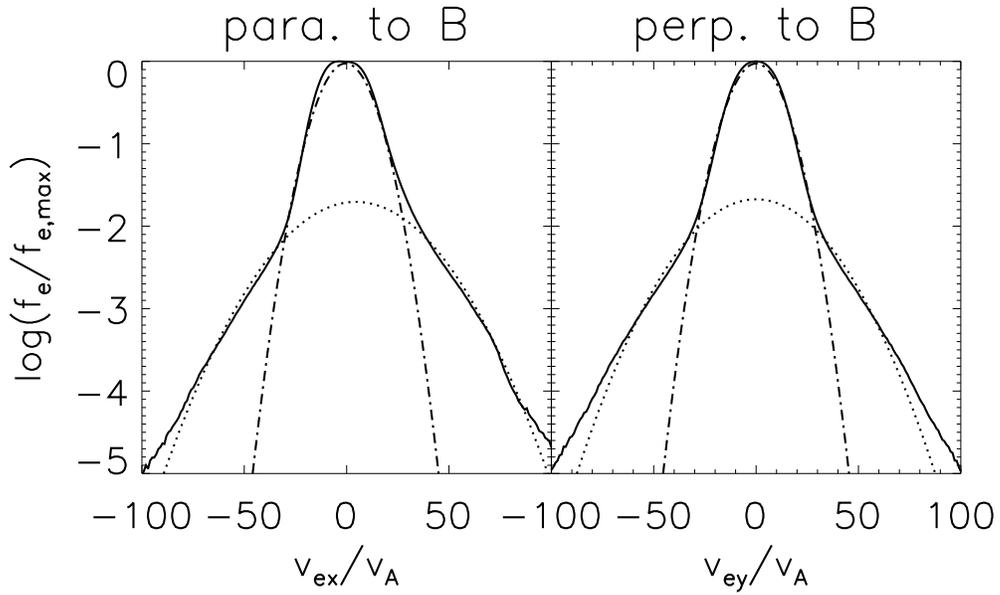}
\caption{ The 1D EVDFs  $f(v_x,v_y=0)$ (parallel to $B_0$) and $f(v_x=0, v_y)$ (perpendicular to $B_0$) at $\Omega_{i} t=10.56$. The dot-dashed lines delineate the core Maxwellian distribution functions with $T_{c}=1 m_i v_A^2$ and the dashed lines represent the halo distribution functions with $T_{h}\sim 6 m_i v_A^2$. }
\label{fhalo}
\end{figure}

In Fig.~\ref{fhalo} we show $f(v_x,v_y=0)$ and $f(v_y,v_x=0)$, cross-sections that cut through the center of the simulated 2D EVDFs  parallel and perpendicular to $B_0$, respectively, after the core-halo feature becomes stable. For theoretical simplicity and to follow the observational convention of \citet{pilipp87jgra,pilipp87jgrb} each of these 1-D EVDFs is fitted with two Maxwellian functions representing the core and the halo respectively, even though the halo distribution is known to be non-Maxwellian. Because both the core and the halo are nearly isotropic, the differences between the two 1-D EVDFs are very small except for the small drift between the halo and core in the $x$-direction. We found the halo temperature $T_{h}$ to be $\sim 6$ times the core temperature $T_{c}$, and the core temperature $T_{c}$ increases by a factor of 4 from the initial temperature $T_{c0}$. The corresponding thermal velocity of the core is $v_{te,c}=\sqrt{k_B T_{c}/m_e}\sim 10 v_A$ which equals the relative core-halo drift and is the condition for saturation of the electron two-stream instability. The ratio between the halo density $n_h$ and the core density $n_c$ is $\sim 0.05$, which is about half the initial density ratio between the electron beam and the ambient electrons.

From what we have learned about the turbulence heating processes from our simulation, we can derive the general properties of core-halo features and relations between these properties.

The EVDF ``break" is simply the velocity at which the core and halo Maxwellians are equal and can be found as
\begin{equation}
v_{brk}\approx[\ln(T_{h}/T_c)-\ln (n_h/n_c)^2]^{1/2} v_{te,c},
\label{brk}
\end{equation}
where we have used the facts that the drift energy of the halo is much smaller than $k_BT_h$ and $1/T_c-1/T_h\sim 1/T_c$. For the parameters we adopted in our simulation, we have $v_{brk}=2.8 v_{te,c}$. 
 
The efficient heating and scattering by kinetic turbulence allows us to determine with relative ease the final temperature of the core $T_{c}$ and $T_h$.  During development of the kinetic turbulence, the total momentum in the center-of-mass frame is zero. The two-stream instability quickly achieves quasi-saturation at $m_e  v_d^2=4 k_B  T_{c\|}$ and the exponential decay ends, where $T_{c\|}$ is the parallel temperature. If the fraction of kinetic energy converted into the heat of  core electrons is $C_T$,  we have  $ n_{c0} k_B T_{c\|}/2-n_{c0} k_B T_{c0}/2 \sim C_T (m_e n_{h0} v_{d0}^2/2-m_e n_{h0} v_{d}^2/2)$, where $v_d$ is the relative drift when the two-stream instability is at its quasi-saturation. Consequently we have 

\begin{equation}
k_B T_{c\|}=\frac{2C_T E_{d0}n_{h0}/n_{c0}+k_B T_{c0}}{1+4C_Tn_{h0}/n_{c0}}.
\label{tcl}
\end{equation}
The perpendicular temperature, on the other hand, is still approximately $T_{c0}$. Later, wave-wave interactions scatter electrons and transfer the thermal momentum from the parallel to perpendicular direction. Part of the beam electrons joins the ambient electrons to form a new core with density $n_c$. Thus the decrease of the new core's parallel thermal energy from $n_c k_B T_{c\|}$ to $n_c k_B T_{c}$ is equal to the increase in perpendicular thermal energy from $n_c k_B  T_{c0}$ to $n_c k_B T_{c}$, which leads to $T_{c\|}=3T_{c}-2T_{c0}$. The final isotropic temperatures  of the core $T_{c}$ and halo $T_{h}$ satisfy the following energy equations (ignoring the small fraction of energy stored in the turbulence):
\begin{equation}
\begin{split}
3n_c k_B (T_c-T_{c0}) + 3n_h k_B (T_h - T_{h0}) =2 n_{h0} E_{d0},
\end{split}
 \label{tc1}
\end{equation}
\begin{equation} 
 3n_c k_B (T_c-T_{c0}) = n_c k_B (T_{c\|}-T_{c0}),
 \label{tc2}  
\end{equation}
where $T_{h0}$ is the initial temperature of beam electrons.

Since $n_{c0}/n_c\approx 1$ and $E_{d0}/k_B T_{c0}\gg 1$, from Eqs.~(\ref{tcl})--(\ref{tc2}) we obtain
\begin{equation}
\frac{T_h-T_{h0}}{T_c-T_{c0}} \approx \frac{n_c}{n_h}\frac{1-C_T}{C_T}+4.
\label{dthc1}
\end{equation}
The final temperatures are much higher than the initial temperatures, then Eq.~(\ref{dthc1}) can be further simplified as 
\begin{equation}
\frac{T_h}{T_c}  \approx \frac{n_c}{n_h}\frac{1-C_T}{C_T}+4.
\label{dthc2}
\end{equation}
Note that  in general $T_h/T_c<(T_h-T_{h0})/(T_c-T_{c0})$ and hence Eq.~(\ref{dthc2}) slightly overestimates $T_h/T_c$. Eq.~(\ref{dthc2}) shows that $T_h/T_c$ depends on the final density ratio $n_h/n_c$ and $C_T$, but is insensitive to the initial condition due to the strong heating process. If $C_t=1$ then $T_h/T_c\sim 4$, meaning that if all of the electron beam kinetic energy is converted into the heat of electrons by the two-stream instability, then the temperature ratio is simply determined by the quasi-saturation condition of the two steam instability, the lower limit of $T_h/T_c$. For a conversion fraction $C_T\sim 0.9$ as shown in our simulation, the temperature ratio varies from $\sim 6-5$ for the reasonable range of $n_h/n_c \in [0.05,0.1]$. 

%\color{blue}
One of the unique predictions of our model is the existence of a relic relative drift between the core and the halo along the magnetic field. The drift equals the core thermal velocity, i.e., $v_{d,relic}=v_{te.c}=\sqrt{k_B T_c/m_e}$ when the two-stream instability fully saturates.

\section{Discussions}
\label{discussion}
In this letter, based on the existing observations and theoretical constraints of nanoflare properties, we show that nanoflare-accelerated electron streams are likely to trigger sufficiently strong electron two-stream instabilities and produce a core-halo feature of EVDF in the inner corona. The core-halo feature can survive the corona collisions and escape to space from the open field lines. We show in Eq.~\ref{dthc2} that the temperature ratio $T_h/T_c$ is related to the quasi-saturation property of the two-stream instability and the density ratio $n_c/n_h$ when the turbulence fully saturates. The ``break'' velocity of the core-halo feature is related to $T_h/T_c$ and $n_c/n_h$ by Eq. (\ref{brk}). A unique feature of our model is a relative drift between core and halo comparable to the core thermal velocity, a relic of the fully saturated electron two-stream instability. 

We now compare the results with the observed EVDFs of slow solar wind with speed $<$~300 \kmps because the slow wind tends to be observed near sector boundaries, where the radial magnetic field is near zero~\citep{erdos12apj}, and thus the magnetic focusing effect is weak and less likely to create distortions to the EVDF as the solar wind travels away from the sun. Features of EVDF such as $T_h/T_c$ and $n_h/n_c$ could be preserved if the core and halo evolve the same way as the solar wind travels into space.  We found that the following observations in the slow wind with speed $<$ 300~\kmps by \citet{pilipp87jgra}  at 1 AU agree well with our model 1) $T_h/T_c\sim 6$; 2) The ``break" is at $v_{\perp} \sim $ 6km/s $\sim 3 v_{te,c}$ (about 100 eV); 3) Pilipp et al. who noted that the relative drifts between the cores and halos in EVDF along the solar wind magnetic field existed, but provided no explanation. We find that the drifts agree well with the thermal velocities of the observed cores in this paper--- a direct consequence of the saturation of the two-stream instability.

The strahl in the solar wind EVDFs is likely produced by magnetic focusing as the solar wind travels outward from the sun. Recent studies suggest that the strahl can form at $10 R_{\bigodot}$ in fast wind~\citep{smith_hm12apj}, well beyond the site where the halo forms. This is consistent with observations showing that strahl originates from both the core and halo~\citep{pilipp87jgra,pilipp87jgrb}. As the solar wind travels to larger heliocentric distances the strahl should become more and more anisotropic. However, observations show that  from 0.3 AU to 1 AU electrons are actually scattered from the strahl into the halo~\citep{mak05jgr,stverak09jgr,gurg12ag}. The cause of the scattering is unclear. The anisotropic strahl may generate new unstable processes such as KAW or whistler wave turbulence when the EVDF is stretched sufficiently~\citep{vocks05apj,rud11pop,mith12pop}. 

During the strahl formation and the strahl-induced electron scattering, the total energy is conserved. Observations of halo and strahl electron densities in the fast wind found that total density of the hot components (halo and strahl) is nearly constant \citep{mak05jgr,stverak09jgr}. The temperature of the hot component electrons $T_{hot}=(n_h T_h + n_s T_s)/(n_h+n_s)$ where $s$ represents the strahl. Then we have $n_{h}T_{h} + n_s T_s=n_{hot} T_{hot}$ where $n_s+n_h = n_{hot}$. Thus Eq. (\ref{dthc2}) can be extended to core-halo-strahl EVDFs, i.e. 
\begin{equation}
\frac{T_{hot}}{T_c} \sim \frac{(1-C_T)}{C_T}\frac{n_c}{ n_{hot}}+4. 
\end{equation}
This ratio is nearly independent to the distance from the sun and even in the fast wind $T_{hot}/T_c \sim 6 $, consistent with the observations~\citep{feldman75jgr}. Further observational test for the evolution of $T_{hot}/T_c$ with the the distance to the sun is needed. 

Nanoflare-accelerated keV electrons are quickly converted into halo electrons in the inner corona. This heating process is so rapid that the keV beam electrons may not produce sufficient non-thermal Xray bremsstrahlung radiation to be observed---consistent with RHESSI observations that found little non-thermal Xray bremsstrahlung radiations at a few keV \citep{hannah10apj}. 

The electron halo may play a very important role in the models of field line launching and accelerating of the solar wind.~\citet{fisk03jgr} proposed a new solar wind model which can produce the observed anti-correlation between the electron temperature and speed of solar wind \citep{gloeckler03jgr}. This model proposed that the magnetized plasma is preheated in the coronal loops and released by reconnection with open magnetic field lines. \citet{gloeckler03jgr} discovered that the electron temperature in corona magnetic loops is higher than that of ambient electrons, confirming the existence of preheating process. However, the nature of this preheating is not well understood. The fast thermalization of electron streams along loops in our model provides a possible explanation for such preheating. 

Electron temperature at the corona source of the solar wind can be obtained from the measurements of charge states of solar wind ions. It has been found that the electron halo in EVDF can affect the ionic charge states and hence ionic charge states can be used to infer properties of electron halo \citep{ko96grl,esser00apjl,laming04apj,feldman08apj}.

\acknowledgements
HC is grateful to the helpful discussions with Drs. J. Klimchunk, B. Dennis, E. Marsch and G. Gloeckler. The authors thank the anomalous referee whose comments helped in making this manuscript more clear and readable. This research was supported by the NASA Postdoctoral Program at NASA/GSFC administered by Oak Ridge Associated Universities and NASA grant NNH11ZDA001N. Resources supporting this work were provided by the NASA High-End Computing (HEC) Program through the NASA Advanced Supercomputing (NAS) Division at Ames Research Center. 

%\bibliography{solarwind}

\begin{thebibliography}{46}
\expandafter\ifx\csname natexlab\endcsname\relax\def\natexlab#1{#1}\fi

\bibitem[{{Aschwanden} \& {Parnell}(2002)}]{aschw02apj}
{Aschwanden}, M.~J. \& {Parnell}, C.~E. 2002, \apj, 572, 1048

\bibitem[{{Che} {et~al.}(2014){Che}, {Goldstein}, \& {Vi{\~n}as}}]{che14prl}
{Che}, H., {Goldstein}, M.~L., \& {Vi{\~n}as}, A.~F. 2014, \prl, 112, 061101

\bibitem[{{Chen} {et~al.}(2013){Chen}, {Bastian}, {White}, {Gary}, {Perley},
  {Rupen}, \& {Carlson}}]{chen_b13apjl}
{Chen}, B., {Bastian}, T.~S., {White}, S.~M., {Gary}, D.~E., {Perley}, R.,
  {Rupen}, M., \& {Carlson}, B. 2013, \apjl, 763, L21

\bibitem[{{Cranmer}(2009)}]{cranmer09lrsp}
{Cranmer}, S.~R. 2009, \lrsp, 6, 3

\bibitem[{{Erd{\H o}s} \& {Balogh}(2012)}]{erdos12apj}
{Erd{\H o}s}, G. \& {Balogh}, A. 2012, \apj, 753, 130

\bibitem[{{Esser} {et~al.}(1995){Esser}, {Brickhouse}, {Habbal}, {Altrock}, \&
  {Hudson}}]{esser95jgr}
{Esser}, R., {Brickhouse}, N.~S., {Habbal}, S.~R., {Altrock}, R.~C., \&
  {Hudson}, H.~S. 1995, \jgr, 100, 19829

\bibitem[{{Esser} \& {Edgar}(2000)}]{esser00apjl}
{Esser}, R. \& {Edgar}, R.~J. 2000, \apjl, 532, L71

\bibitem[{{Feldman} {et~al.}(2005){Feldman}, {Landi}, \&
  {Schwadron}}]{feldman05jgr}
{Feldman}, U., {Landi}, E., \& {Schwadron}, N.~A. 2005, \jgr, 110, 7109

\bibitem[{{Feldman} {et~al.}(2008){Feldman}, {Ralchenko}, \&
  {Landi}}]{feldman08apj}
{Feldman}, U., {Ralchenko}, Y., \& {Landi}, E. 2008, \apj, 684, 707

\bibitem[{{Feldman} {et~al.}(1999){Feldman}, {Widing}, \&
  {Warren}}]{feldman99apj}
{Feldman}, U., {Widing}, K.~G., \& {Warren}, H.~P. 1999, \apj, 522, 1133

\bibitem[{{Feldman} {et~al.}(1975){Feldman}, {Asbridge}, {Bame}, {Montgomery},
  \& {Gary}}]{feldman75jgr}
{Feldman}, W.~C., {Asbridge}, J.~R., {Bame}, S.~J., {Montgomery}, M.~D., \&
  {Gary}, S.~P. 1975, \jgr, 80, 4181

\bibitem[{{Fisk}(2003)}]{fisk03jgr}
{Fisk}, L.~A. 2003, \jgr, 108, 1157

\bibitem[{{Gloeckler} {et~al.}(2003){Gloeckler}, {Zurbuchen}, \&
  {Geiss}}]{gloeckler03jgr}
{Gloeckler}, G., {Zurbuchen}, T.~H., \& {Geiss}, J. 2003, \jgr, 108, 1158

\bibitem[{{Gontikakis} {et~al.}(2013){Gontikakis}, {Patsourakos},
  {Efthymiopoulos}, {Anastasiadis}, \& {Georgoulis}}]{gon13apj}
{Gontikakis}, C., {Patsourakos}, S., {Efthymiopoulos}, C., {Anastasiadis}, A.,
  \& {Georgoulis}, M.~K. 2013, \apj, 771, 126

\bibitem[{{Gopalswamy} {et~al.}(2012){Gopalswamy}, {Nitta}, {Akiyama},
  {M{\"a}kel{\"a}}, \& {Yashiro}}]{gopa12apj}
{Gopalswamy}, N., {Nitta}, N., {Akiyama}, S., {M{\"a}kel{\"a}}, P., \&
  {Yashiro}, S. 2012, \apj, 744, 72

\bibitem[{{Gurgiolo} {et~al.}(2012){Gurgiolo}, {Goldstein}, {Vi{\~n}as}, \&
  {Fazakerley}}]{gurg12ag}
{Gurgiolo}, C., {Goldstein}, M.~L., {Vi{\~n}as}, A.~F., \& {Fazakerley}, A.~N.
  2012, \ag, 30, 163

\bibitem[{{Hannah} {et~al.}(2010){Hannah}, {Hudson}, {Hurford}, \&
  {Lin}}]{hannah10apj}
{Hannah}, I.~G., {Hudson}, H.~S., {Hurford}, G.~J., \& {Lin}, R.~P. 2010, \apj,
  724, 487

\bibitem[{{He} {et~al.}(2010){He}, {Tu}, {Tian}, \& {Marsch}}]{he10asr}
{He}, J.-S., {Tu}, C.-Y., {Tian}, H., \& {Marsch}, E. 2010, \asr, 45, 303

\bibitem[{{Jockers}(1970)}]{jock70aap}
{Jockers}, K. 1970, \aap, 6, 219

\bibitem[{{Klimchuk}(2012)}]{klim12jgr}
{Klimchuk}, J.~A. 2012, \jgr, 117, 12102

\bibitem[{{Ko} {et~al.}(1996){Ko}, {Fisk}, {Gloeckler}, \& {Geiss}}]{ko96grl}
{Ko}, Y.-K., {Fisk}, L.~A., {Gloeckler}, G., \& {Geiss}, J. 1996, \grl, 23,
  2785

\bibitem[{{Krucker} {et~al.}(2010){Krucker}, {Hudson}, {Glesener}, {White},
  {Masuda}, {Wuelser}, \& {Lin}}]{krucker10apj}
{Krucker}, S., {Hudson}, H.~S., {Glesener}, L., {White}, S.~M., {Masuda}, S.,
  {Wuelser}, J.-P., \& {Lin}, R.~P. 2010, \apj, 714, 1108

\bibitem[{{Laming}(2004)}]{laming04apj}
{Laming}, J.~M. 2004, \apj, 604, 874

\bibitem[{{Lin}(1997)}]{lin97conf}
{Lin}, R.~P. 1997, in AIP Conference Proceedings, Vol. 385, Scientific Basis
  for Robotic Exploration Close to the Sun, AIP (American Institute of
  Physics), 25--32

\bibitem[{{Lin}(2011)}]{lin11ssr}
{Lin}, R.~P. 2011, \ssr, 159, 421

\bibitem[{{Maksimovic} {et~al.}(2005){Maksimovic}, {Zouganelis}, {Chaufray},
  {Issautier}, {Scime}, {Littleton}, {Marsch}, {McComas}, {Salem}, {Lin}, \&
  {Elliott}}]{mak05jgr}
{Maksimovic}, M., {Zouganelis}, I., {Chaufray}, J.-Y., {Issautier}, K.,
  {Scime}, E.~E., {Littleton}, J.~E., {Marsch}, E., {McComas}, D.~J., {Salem},
  C., {Lin}, R.~P., \& {Elliott}, H. 2005, \jgr, 110, 9104

\bibitem[{{Marsch}(2006)}]{marsch06lrsp}
{Marsch}, E. 2006, \lrsp, 3, 1

\bibitem[{{Mithaiwala} {et~al.}(2012){Mithaiwala}, {Rudakov}, {Crabtree}, \&
  {Ganguli}}]{mith12pop}
{Mithaiwala}, M., {Rudakov}, L., {Crabtree}, C., \& {Ganguli}, G. 2012, \pop,
  19, 102902

\bibitem[{{Parker}(1965)}]{parker65ssr}
{Parker}, E.~N. 1965, \ssr, 4, 666

\bibitem[{{Parker}(1988)}]{parker88apj}
---. 1988, \apj, 330, 474

\bibitem[{{Pilipp} {et~al.}(1987{\natexlab{a}}){Pilipp}, {Muehlhaeuser},
  {Miggenrieder}, {Montgomery}, \& {Rosenbauer}}]{pilipp87jgra}
{Pilipp}, W.~G., {Muehlhaeuser}, K.-H., {Miggenrieder}, H., {Montgomery},
  M.~D., \& {Rosenbauer}, H. 1987{\natexlab{a}}, \jgr, 92, 1075

\bibitem[{{Pilipp} {et~al.}(1987{\natexlab{b}}){Pilipp}, {Muehlhaeuser},
  {Miggenrieder}, {Rosenbauer}, \& {Schwenn}}]{pilipp87jgrb}
{Pilipp}, W.~G., {Muehlhaeuser}, K.-H., {Miggenrieder}, H., {Rosenbauer}, H.,
  \& {Schwenn}, R. 1987{\natexlab{b}}, \jgr, 92, 1103

\bibitem[{{Rudakov} {et~al.}(2011){Rudakov}, {Mithaiwala}, {Ganguli}, \&
  {Crabtree}}]{rud11pop}
{Rudakov}, L., {Mithaiwala}, M., {Ganguli}, G., \& {Crabtree}, C. 2011, \pop,
  18, 012307

\bibitem[{{Scudder}(1992{\natexlab{a}})}]{scudder92apja}
{Scudder}, J.~D. 1992{\natexlab{a}}, \apj, 398, 299

\bibitem[{{Scudder}(1992{\natexlab{b}})}]{scudder92apjb}
---. 1992{\natexlab{b}}, \apj, 398, 319

\bibitem[{{Smith} {et~al.}(2012){Smith}, {Marsch}, \&
  {Helander}}]{smith_hm12apj}
{Smith}, H.~M., {Marsch}, E., \& {Helander}, P. 2012, \apj, 753, 31

\bibitem[{{Testa} {et~al.}(2013){Testa}, {De Pontieu},
  {Mart{\'{\i}}nez-Sykora}, {DeLuca}, {Hansteen}, {Cirtain}, {Winebarger},
  {Golub}, {Kobayashi}, {Korreck}, {Kuzin}, {Walsh}, {DeForest}, {Title}, \&
  {Weber}}]{testa13apjl}
{Testa}, P., {De Pontieu}, B., {Mart{\'{\i}}nez-Sykora}, J., {DeLuca}, E.,
  {Hansteen}, V., {Cirtain}, J., {Winebarger}, A., {Golub}, L., {Kobayashi},
  K., {Korreck}, K., {Kuzin}, S., {Walsh}, R., {DeForest}, C., {Title}, A., \&
  {Weber}, M. 2013, \apjl, 770, L1

\bibitem[{{{\v S}tver{\'a}k} {et~al.}(2009){{\v S}tver{\'a}k}, {Maksimovic},
  {Tr{\'a}vn{\'{\i}}{\v c}ek}, {Marsch}, {Fazakerley}, \&
  {Scime}}]{stverak09jgr}
{{\v S}tver{\'a}k}, {\v S}., {Maksimovic}, M., {Tr{\'a}vn{\'{\i}}{\v c}ek},
  P.~M., {Marsch}, E., {Fazakerley}, A.~N., \& {Scime}, E.~E. 2009, \jgr, 114,
  5104

\bibitem[{{van der Holst} {et~al.}(2010){van der Holst}, {Manchester},
  {Frazin}, {V{\'a}squez}, {T{\'o}th}, \& {Gombosi}}]{van10apj}
{van der Holst}, B., {Manchester}, IV, W.~B., {Frazin}, R.~A., {V{\'a}squez},
  A.~M., {T{\'o}th}, G., \& {Gombosi}, T.~I. 2010, \apj, 725, 1373

\bibitem[{{Vekstein}(2009)}]{vekstein09aap}
{Vekstein}, G. 2009, \aap, 499, L5

\bibitem[{{Viall} \& {Klimchuk}(2013{\natexlab{a}})}]{viall13book}
{Viall}, N. \& {Klimchuk}, J.~A. 2013{\natexlab{a}}, in AAS Solar Physics
  Division Meeting, Vol.~44, AAS Solar Physics Division Meeting, \#{\ldots}16

\bibitem[{{Viall} \& {Klimchuk}(2013{\natexlab{b}})}]{viall13apj}
{Viall}, N.~M. \& {Klimchuk}, J.~A. 2013{\natexlab{b}}, \apj, 771, 115

\bibitem[{{Vocks}(2012)}]{vocks12ssr}
{Vocks}, C. 2012, \ssr, 172, 303

\bibitem[{{Vocks} {et~al.}(2005){Vocks}, {Salem}, {Lin}, \&
  {Mann}}]{vocks05apj}
{Vocks}, C., {Salem}, C., {Lin}, R.~P., \& {Mann}, G. 2005, \apj, 627, 540

\bibitem[{{Wang} {et~al.}(2012){Wang}, {Lin}, {Salem}, {Pulupa}, {Larson},
  {Yoon}, \& {Luhmann}}]{wang12apjl}
{Wang}, L., {Lin}, R.~P., {Salem}, C., {Pulupa}, M., {Larson}, D.~E., {Yoon},
  P.~H., \& {Luhmann}, J.~G. 2012, \apjl, 753, L23

\bibitem[{{Winebarger} {et~al.}(2013){Winebarger}, {Walsh}, {Moore}, {De
  Pontieu}, {Hansteen}, {Cirtain}, {Golub}, {Kobayashi}, {Korreck}, {DeForest},
  {Weber}, {Title}, \& {Kuzin}}]{win13apj}
{Winebarger}, A.~R., {Walsh}, R.~W., {Moore}, R., {De Pontieu}, B., {Hansteen},
  V., {Cirtain}, J., {Golub}, L., {Kobayashi}, K., {Korreck}, K., {DeForest},
  C., {Weber}, M., {Title}, A., \& {Kuzin}, S. 2013, \apj, 771, 21

\end{thebibliography}

\vfil\eject\clearpage

%Fig 1==========
\newpage
%Fig 2============

\end{document}